\documentclass[%
 reprint,
 amsmath,amssymb,
 aps,
]{revtex4-2}

\usepackage{graphicx}
\usepackage{dcolumn}
\usepackage{bm}
\usepackage{comment}
\usepackage{dsfont}
\usepackage{xcolor}

\begin{document}

\preprint{APS/123-QED}

\title{Non-Markovian protection of states from decay in quasi-PT-symmetric systems}

\author{Timofey T. Sergeev}
\affiliation{Moscow Institute of Physics and Technology, 141700, 9 Institutskiy pereulok, Moscow, Russia}
\affiliation{Dukhov Research Institute of Automatics (VNIIA), 127055, 22 Sushchevskaya, Moscow, Russia}
\affiliation{Institute for Theoretical and Applied Electromagnetics, 125412, 13 Izhorskaya, Moscow, Russia}
\affiliation{Institute of Spectroscopy Russian Academy of Sciences, 108840, 5 Fizicheskaya, Troitsk, Moscow, Russia}
\author{Evgeny S. Andrianov}
\affiliation{Moscow Institute of Physics and Technology, 141700, 9 Institutskiy pereulok, Moscow, Russia}
\affiliation{Dukhov Research Institute of Automatics (VNIIA), 127055, 22 Sushchevskaya, Moscow, Russia}
\affiliation{Institute for Theoretical and Applied Electromagnetics, 125412, 13 Izhorskaya, Moscow, Russia}
\author{Alexander A. Zyablovsky}
\email{zyablovskiy@mail.ru}
\affiliation{Moscow Institute of Physics and Technology, 141700, 9 Institutskiy pereulok, Moscow, Russia}
\affiliation{Dukhov Research Institute of Automatics (VNIIA), 127055, 22 Sushchevskaya, Moscow, Russia}
\affiliation{Institute for Theoretical and Applied Electromagnetics, 125412, 13 Izhorskaya, Moscow, Russia}

\date{\today}

\begin{abstract}
   We consider a quasi-PT-symmetric system of two resonators, one of which interacts with a finite-size environment. The interaction with the environment leads to energy losses in the resonators, and the finite size of the environment leads to a non-Markovian dynamics of the relaxation process. We demonstrate that non-Markovian processes in the quasi-PT-symmetric system can make the states of the system infinitely living, loss-protected states, even in the absence of gain. There is a critical value of the interaction between the resonator and the environment below which any state of the system is loss-protected. When the interaction magnitude is greater than the critical value, depending on the coupling strength between the resonators, either one or both states are unprotected. We show that the boundaries of regions with different numbers of protected states are determined by the relaxation rates in the quasi-PT-symmetric system, calculated in the Markovian approximation. By changing the coupling strength between the resonators and the interaction magnitude between the resonator and the environment, the system switches between modes with two, one, or no loss-protected states. This makes it possible to realize stable PT-symmetric devices based on purely dissipative systems. The obtained results are applicable to quantum systems with single excitations, allowing the concept of PT symmetry to be extended to such systems. 
\end{abstract}

\maketitle

\textit{Introduction.} Non-Hermitian systems with exceptional points are objects of a comprehensive study \cite{5,6,11,12}. Exceptional points are singularities in the system parameter space in which several eigenstates are linearly dependent, and their eigenvalues coincide with each other \cite{5,6,11,12}. By changing the system parameters, it is possible to make a transition through an exceptional point \cite{5,6,11,12,21,22}. This transition leads to qualitative changes in the eigenstates and eigenvalues of the system \cite{5,6,11,12}. PT-symmetric systems that combine absorbing and amplifying elements are an example of non-Hermitian systems with exceptional points \cite{13,14,15,16,17,18}. In these systems, the transition at the exceptional point is accompanied by spontaneous PT-symmetry breaking for the eigenstates \cite{13,14,15,16,17,18}. In the PT-symmetrical phase, the eigenstates have the same distribution in the absorbing and amplifying elements, which leads to a precise compensation of losses \cite{13,14,15,16,17,18}. In the non-PT-symmetrical phase, the eigenstates are amplifying or absorbing \cite{13,14,15,16,17,18}.

The exact balance between amplification and absorption is difficult to achieve in practice. Therefore, one usually deals with quasi-PT-symmetric systems \cite{19,23,24,25,20}, which do not provide accurate loss compensation but exhibit transitions with spontaneous PT-symmetry breaking \cite{19,20,23,24,25}. The simplest example of a quasi-PT-symmetric system is a structure of two coupled waveguides, the first of which is non-dissipative and the second is dissipative \cite{19,20}. In such a system, there is an exceptional point that occurs at the critical value of the coupling strength between the waveguide. When the coupling strength between the waveguides is less than the critical value, one of the eigenmodes has a maximum in the first waveguide and a small relaxation rate. The second eigenmode has a maximum in the second waveguide and a larger relaxation rate. When the coupling strength is greater than the critical value, the eigenmodes have the same amplitudes in the two waveguides and equal relaxation rates.

In recent years, the influence of non-Markov processes on the behavior of non-Hermitian systems with exceptional points has become the object of close study \cite{8,27,28,29,30,31,32}. Non-Markovian processes provide an additional degree of control over the behavior of non-Hermitian systems. In particular, non-Markovian effects can lead to the emergence of additional or higher-order exceptional points \cite{27}, which can be used, for example, in metrology.

In this letter, we demonstrate that non-Hermitian processes can protect non-Hermitian systems from dissipation, leading to the formation of infinitely long-lived states even in the absence of amplification. We consider a system of two resonators, one of which interacts with a finite-size environment. The interaction with the reservoir leads to energy losses in the resonators. In the limit of an infinite environment, the interaction of the system with the environment can be described by introducing a relaxation term into the equations. In this approximation, the system is quasi-PT-symmetric and exhibits a transition at an exceptional point. The finite size of the environment leads to a non-Markovian dynamics of the relaxation process when interaction with the environment is determined not only by the current, but also by previous moments in time. We demonstrate that non-Markovian processes in such a quasi-PT-symmetric system can make the state of the system infinitely living, state loss-protected by non-Markovian effects, even in the absence of amplification. We show that there is a critical value in the magnitude of the interaction between the resonator and the environment. When the magnitude of the interaction, $g$, is less than the critical value, any state of the system is loss-protected. When the interaction is greater than the critical value, depending on the coupling strength between the resonators, $\Omega$, either only one state is loss-protected, or there are no protected states at all. At small values of $\Omega$, one state is loss-protected and the other is unprotected. With increasing $\Omega$, all states cease to be loss-protected. We demonstrate that the boundaries of regions with different numbers of loss-protected states are determined by the relaxation rates in the quasi-PT-symmetric system. In particular, the boundaries of the regions clearly feel the transition through the exceptional point in the quasi-PT-symmetric system. This result is applicable to both classical and quantum systems, which opens the way to efficient control of quantum states for quantum information processing tasks.

\textit{Model.} We consider a quantum system consisting of two coupled single-mode resonators, where one of them interacts with a finite-sized environment. We consider the frequencies of single-mode resonators to coincide and to be equal to $\omega_0$. We describe the environment as a finite set of $N$ modes with an equidistant distribution of frequencies that lie close to $\omega_0$ and have the form $\omega_j = \omega_0 + \delta\omega(j-N/2)$, where $\delta\omega$ is a step between the frequencies of the modes. 

To describe the system, we use the following Hamiltonian \cite{1}:

\begin{equation}
\begin{array}{l}
\hat H = {\omega _0}\hat a_1^\dag {{\hat a_1}} + {\omega _0}\hat a_2^\dag {{\hat a_2}} + \Omega (\hat a_1 {{\hat a_2^{\dag} }} + \hat a_1^{\dag} {{\hat a_2}}) + \\
\sum\limits_{j = 1}^{N} {{\omega_j \hat b_j^\dag {{\hat b}_j}}} + \sum\limits_{j = 1}^{N} {{g_j}(\hat a_2 {{\hat b}_j^{\dag}} + \hat a_2^{\dag} {{\hat b}_j})}
\end{array}
\label{eq:1}
\end{equation}
where $\hat a_{1,2}$ and $\hat a_{1,2}^{\dag}$ are the annihilation and creation operators of two single-mode resonators that obey the bosonic commutation relation $[\hat a_{1,2},\hat a_{1,2}^{\dag}] = 1$. $\hat b_j$, $\hat b_j^{\dag}$ are the annihilation and creation operators of the environment modes that also obey the bosonic commutation relation $[\hat b_i, \hat b_j^{\dag}] = \delta_{ij}$. $\Omega$ is a coupling strength between the first and second single-mode cavities. $g_j$ is a coupling strength between the second single-mode cavities and the $j$-th mode of the environment. In the following, we consider the environment modes to interact with the second resonator with the same coupling strength, i.e. $g_j = g$ for all $j$. $N$ is a number of environment modes.

To describe the dynamics of the system, we use the time-dependent Schr\"{o}dinger equations \cite{26}, where we look for the wave function in the following form:

\begin{equation}
\vert\Psi(t)\rangle =  a_1(t) \vert1,0,0\rangle + a_2(t) \vert0,1,0\rangle + \sum\limits_{j=1}^{N} {b_j(t) \vert0,0,1_j\rangle} 
\label{eq:2}
\end{equation}
where $a_1(t)$, $a_2(t)$ and $b_j(t)$ are amplitudes of probability of excitation quantum to be in the first and second single-mode cavities, or in one of the modes of the environment, respectively. After substituting the wave function (\ref{eq:2}) into the Schr\"{o}dinger equation, we can obtain a closed system of equations for amplitudes $a_{1,2}$ and $b_j$.:

\begin{equation}
\frac{{d{a_1}}}{{dt}} =  - i{\omega _0}{a_1} - i\,\Omega {a_2} 
\label{eq:3}
\end{equation}

\begin{equation}
\frac{{d{a_2}}}{{dt}} =  - i{\omega _0}{a_2} - i\,\Omega {a_1}-\sum\limits_{j =1}^{N} {i g{b_j}}
\label{eq:4}
\end{equation}

\begin{equation}
\frac{{d{b_j}}}{{dt}} =  - i{\omega_j}{b_j} - ig{a_2}
\label{eq:5}
\end{equation}

The equations~(\ref{eq:3})-(\ref{eq:5}) describe the dynamics of the probability amplitudes of the resonator and surrounding modes. The same equations are obtained for the mode amplitudes in the classical limit, when the number of excitations in the modes is much greater than 1. Therefore, the obtained results are applicable to both quantum and classical systems.

\textit{Limit of infinite environment.} The description of the system of equations~(\ref{eq:3})-(\ref{eq:5}) can be simplified within the Born-Markovian approximation \cite{1,2,3}. In such a description, we can eliminate the environment's modes degrees of freedom and obtain the non-Hermitian system of equations \cite{4,8,33,34}:

\begin{equation}
\frac{d}{{dt}}\left( {\begin{array}{*{20}{c}}
{{a_1}}\\
{{a_2}}
\end{array}} \right) = \left( {\begin{array}{*{20}{c}}
{ - i{\omega _0}}&{ - i\Omega }\\
{ - i\Omega }&{ - i{\omega _0} - \gamma}
\end{array}} \right)\left( {\begin{array}{*{20}{c}}
{{a_1}}\\
{{a_2}}
\end{array}} \right)
\label{eq:6}
\end{equation}
where $\gamma=\pi g^2/\delta\omega$ is an effective decay rate, obtained through elimination of environment's degrees of freedom within the Born-Markov approximation \cite{1,2,3,4}. 

Non-Hermitian systems are famous for the presence of an exceptional point (EP), where the eigenvalues of the system become equal to each other and eigenvectors become collinear \cite{4,5,6,7,8}. The eigenvalues and eigenvectors of the equations~(\ref{eq:6}) are determined by the following expressions:

\begin{equation}
\lambda_{\pm}=-\frac{\gamma}{2} -i\omega_0 \pm \frac{1}{2}\sqrt{\gamma^2 - 4 \Omega^2}
\label{eq:7}
\end{equation}

\begin{equation}
\textbf{e}_\pm= \{i(\gamma\pm\sqrt{\gamma^2-4\Omega^2})/2\Omega, \quad1\}^T
\label{eq:8}
\end{equation}

Here, there is an exceptional point when $\Omega=\Omega_{EP}=\gamma/2$. At the EP the eigenvalues~(\ref{eq:7}) coincide and eigenvectors~(\ref{eq:8}) are collinear. Passing through the exceptional point is often associated with non-Hermitian phase transition \cite{5,6,7,8} that also accompanied by the spontaneous symmetry breaking in system's eigenstates \cite{5,6,7}. When $\Omega < \Omega_{EP}$, the eigenvectors~(\ref{eq:8}) are PT-symmetrical and the real part of eigenvalues~(\ref{eq:7}) are equal to each other. That is, the eigenstates have the same relaxation rates, $\Gamma_{1,2}$, which equal to $-\gamma /2$; t. When $\Omega>\Omega_{EP}$, the eigenvectors~(\ref{eq:8}) are non-PT-symmetrical and the real part of the eigenvalues are different from each other ($\Gamma_1 \ne \Gamma_2$). For all values of $\Omega$ except $\Omega=0$, the real parts of both eigenvalues are negative, that is, the eigenstates are damped ($\Gamma_{1,2} \ne 0$).

\begin{figure}[htbp]
\centering\includegraphics[width=0.8\linewidth]{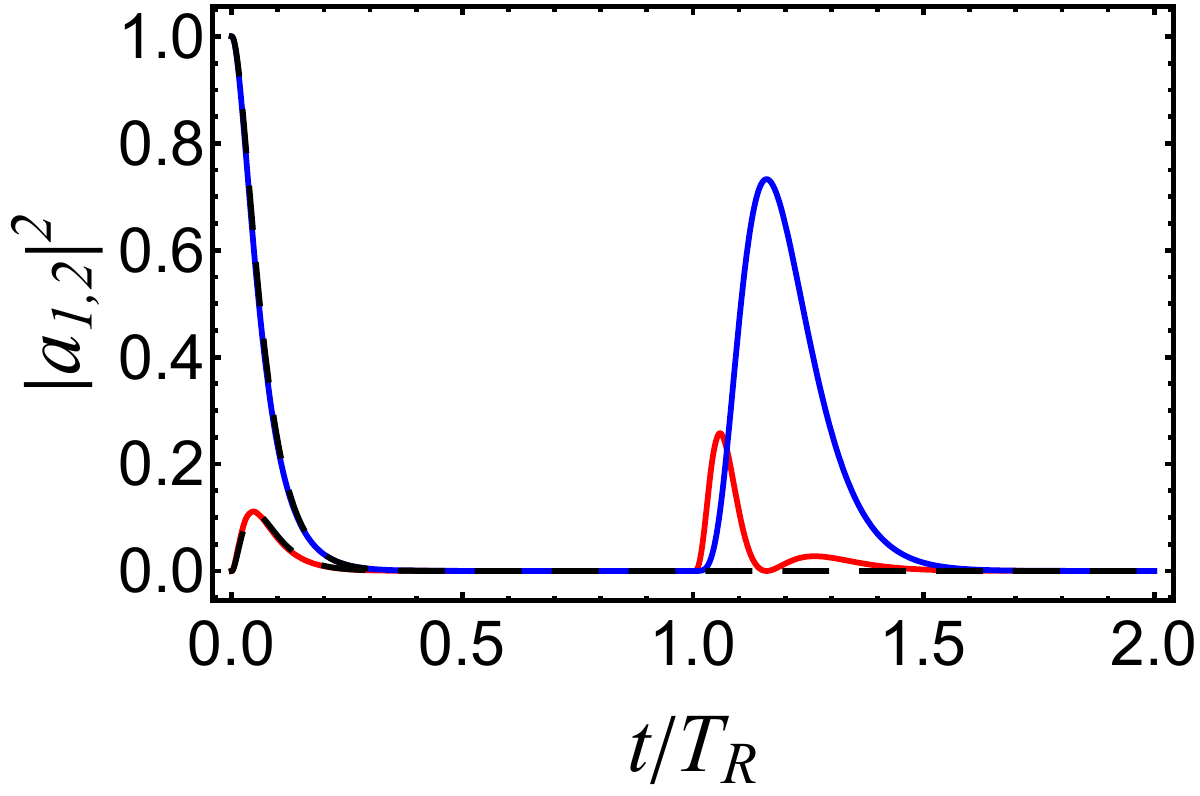}
\caption{Dependence of the first (blue solid line) and second (red solid line) resonator's probability amplitudes $\vert a_{1,2}(t)\vert^{2}$ on time calculated by the Eqns.~(\ref{eq:3})-(\ref{eq:5}) in the case $\gamma>>\delta\omega$ and $\Omega \sim \gamma$. The black dashed lines corresponds to the $\vert a_{1,2}(t)\vert^{2}$ calculated by non-Hermitian equations~(\ref{eq:6}). Here, $T_R=2\pi/\delta\omega$ is the time of first revival. We consider $N=100$, $\delta\omega = 2 \times 10^{-3} \omega_0$, $g=3\times 10^{-3} \omega_0$, $\gamma=\pi g^2/\delta\omega \approx 1.4\times 10^{-2} \omega_0>>\delta\omega$, $\Omega=6 \times 10^{-3} \omega_0$.}
\label{fig1}
\end{figure}

\textit{The case of an environment of finite size.} The non-Hermitian equation~(\ref{eq:6}) correctly describes the dynamics of system~(\ref{eq:3})-(\ref{eq:5}) only in the limit of an infinite environment. The finite size of the environment leads to non-Markovian effects. When the energy flow from the resonators leads to the excitation of the environment modes, which in turn leads to the formation of the reverse flow from the environment to the resonators. The reverse energy flow leads to a phenomenon such as the revival of oscillations in the resonators \cite{10}. At times multiples of $t=T_R=2\pi/\delta\omega$, a sharp increase in the amplitude of oscillations in the resonators is observed \cite{9,10} [see Figure \ref{fig1}]. As a result, the exponential decay of the amplitudes of the resonator modes predicted by non-Hermitian equations~(\ref{eq:6}) is observed only when $t<T_R$. At large times, the backflow of energy from the environment maintains the oscillations in the resonator modes [Figure~\ref{fig1}].

\begin{figure}[htbp]
\centering\includegraphics[width=0.8\linewidth]{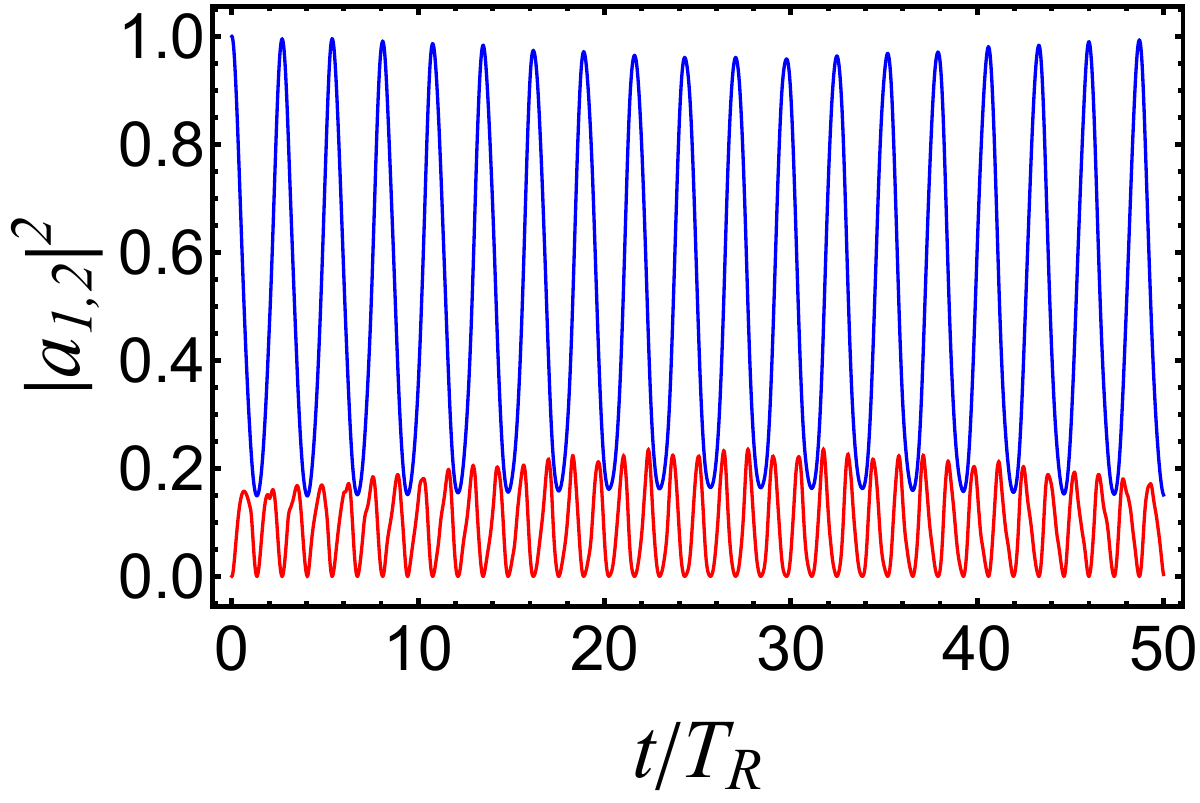}
\caption{Dependence of the first (blue solid line) and second (red solid line) resonator's probability amplitudes $\vert a_{1,2}(t)\vert^{2}$ on time calculated by the Eqns.~(\ref{eq:3})-(\ref{eq:5}) in the case $\gamma \lesssim \delta\omega$. Here, $T_R=2\pi/\delta\omega$ is the time of first revival. We consider $N=100$, $\delta\omega = 2 \times 10^{-3} \omega_0$, $g=7.5 \times 10^{-4} \omega_0$, $\gamma=\pi g^2/\delta\omega \approx 8.8\times 10^{-4} \omega_0<\delta\omega$, $\Omega=5 \times 10^{-4} \omega_0$.}
\label{fig2}
\end{figure}

To describe the system evolution at $t>T_R$, we have to consider a Hermitian system of equations~(\ref{eq:3})-(\ref{eq:5}). The dynamics of the system strongly depend on $\Omega$, $g$ and the initial states. Our calculations show that the backflow of energy from the environment to the resonator can prevent the decay of the oscillations in the resonator modes, making the state of the system infinitely living [Figure~\ref{fig2}].
In what follows, we call such infinitely long-lived states loss-protected ones. There are regions of parameters in which all states of the system are loss-protected, one specific state is loss-protected, or there are no protected states.

To characterize the time of life of the states, we use a memory, $M$, which is defined by the following expression:

\begin{equation}
M=\frac{1}{T}\int\limits_{\tau}^{\tau+T}dt \vert\langle \Psi(0) \vert \Psi(t)\rangle\vert^2
\label{eq:10}
\end{equation}
where $\tau$ and $T$ are much greater than $T_R$. The memory is the average value of a quantity $\vert\langle \Psi(0) \vert \Psi(t)\rangle\vert^2$ that shows the probability that at a moment in time $t$ the state of the system coincides with its initial state. If, for arbitrarily large values of $\tau$ and $T$, the memory, $M$, is close to $1$, then the chosen initial state is infinitely living (loss-protected).

\begin{figure}[htbp]
\centering\includegraphics[width=\linewidth]{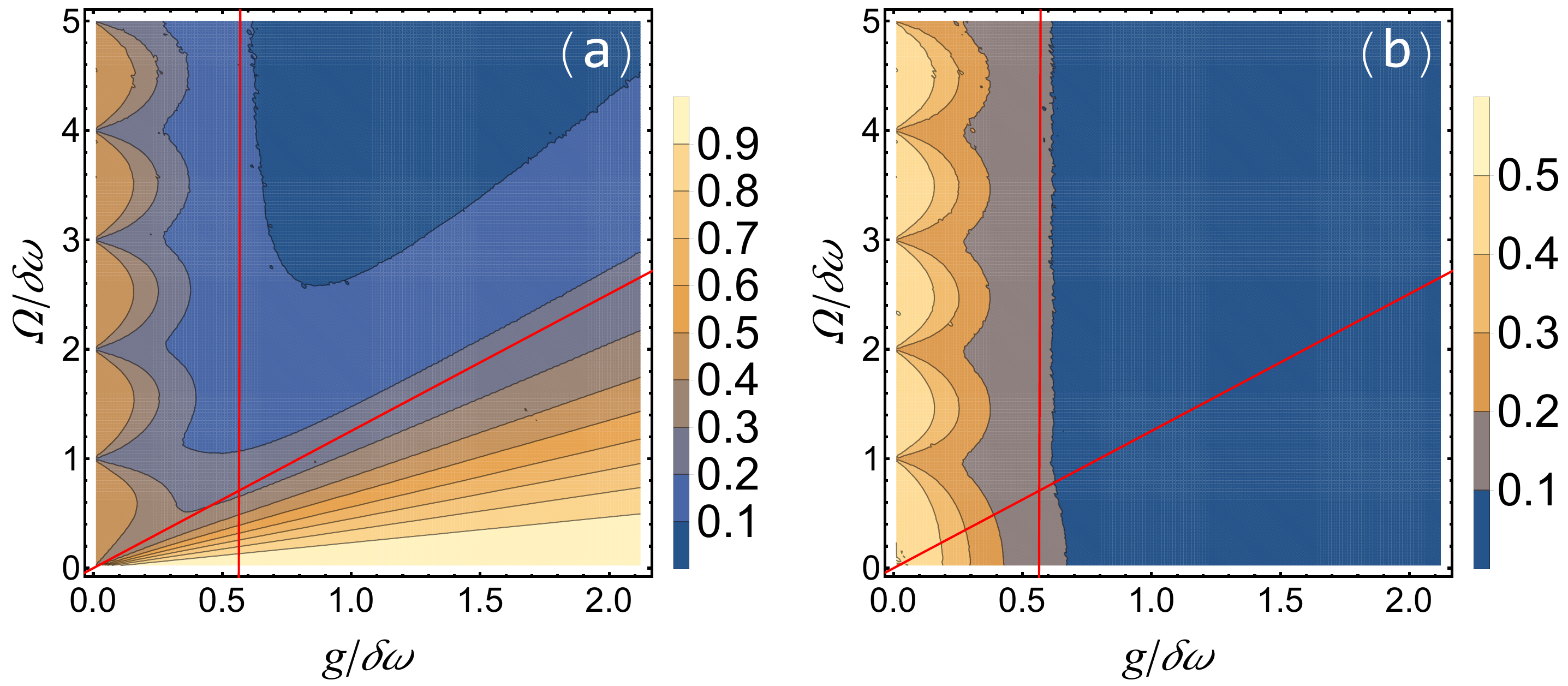}
\caption{Dependence of the system's memory $M$ on the coupling strengths $g$ and $\Omega$ with the initial condition $a_1(t=0)=1$, $a_2(t=0)=0$ (a) and $a_1(t=0)=0$, $a_2(t=0)=1$ (b). Here, $N=50$, $\delta\omega = 2 \times 10^{-3} \omega_0$. Vertical red line corresponds to the condition $\gamma =\delta\omega$ (or $g=\delta\omega/\sqrt{\pi}$). The inclined red line corresponds to the condition $\Gamma_1\ \approx 2\Omega^2/\gamma =\delta\omega$ (or $\Omega=\sqrt{\frac{\pi}{2}}g$).}
\label{fig3}
\end{figure}

We calculate the dependence of memory on the coupling strengths $g$ and $\Omega$ for different initial states. Figure \ref{fig3} shows that there is a parameter range ($\gamma\sim g^2/\delta\omega \lesssim \delta\omega$) where both initial states $a_1(t=0)=1$, $a_2(t=0)=0$; and $a_1(t=0)=0$, $a_2(t=0)=1$ give a dynamics that preserves the memory of the initial state. That is, both states are loss-protected, and due to the linearity of the problem, any state is loss-protected. There is a parameter range ($g>\delta\omega$ and $\Gamma_1\ \approx 2\Omega^2/\gamma \lesssim\delta\omega$) where only state $a_1(t=0)=1$ and $a_2(t=0)=0$ give dynamics with memory retention. In this case, only the state in which the oscillations are excited in the resonator that does not interact with the environment is loss-protected. In the parameter range where $g>\delta\omega$ and $\Omega \gtrsim \gamma$, no initial state gives dynamics with memory retention. Thus, depending on the coupling strengths $g$ and $\Omega$, the number of loss-protected states can vary from two to zero. By changing the parameters of the system, we can change the number of protected states, which can be useful for controlling quantum states in computing devices.

\textit{The influence of the PT-symmetry transition on the time of life of states.} The change in the time of life of states with a change of $g$ and $\Omega$ is associated with the PT-transition in the quasi-PT-symmetric system. At $t<T_R$, the energy flow from the resonators to the environment modes. This flow leads to exponential decay of the amplitudes of the resonator modes and excitation of the environment modes. The modes whose frequencies lie within the radiation line width are predominantly excited. Since in the initial stage the evolution of the Hermitian system~(\ref{eq:3})-(\ref{eq:5}) coincides with the evolution of the non-Hermitian system~(\ref{eq:6}), the radiation line is determined by the relaxation rates of the eigenmodes~(\ref{eq:7}). The number of environmental modes excited by the energy flow from the resonators can be estimated as the ratio of the relaxation rates to the step between the frequencies of the modes: $N_{ex} \sim \Gamma_{1,2} / \delta \omega$ (the relaxation rate and the number of excited modes depend on the initial conditions). If $N_{ex} >> 1$ then the initial excitation is distributed over a large number of modes and at large times the state of the system turns out to be very different from the initial state. That is, the states are not loss-protected. The initial state is loss-protected if the number of excited modes $N_{ex} \lesssim 1$.

\begin{figure}[htbp]
\centering\includegraphics[width=\linewidth]{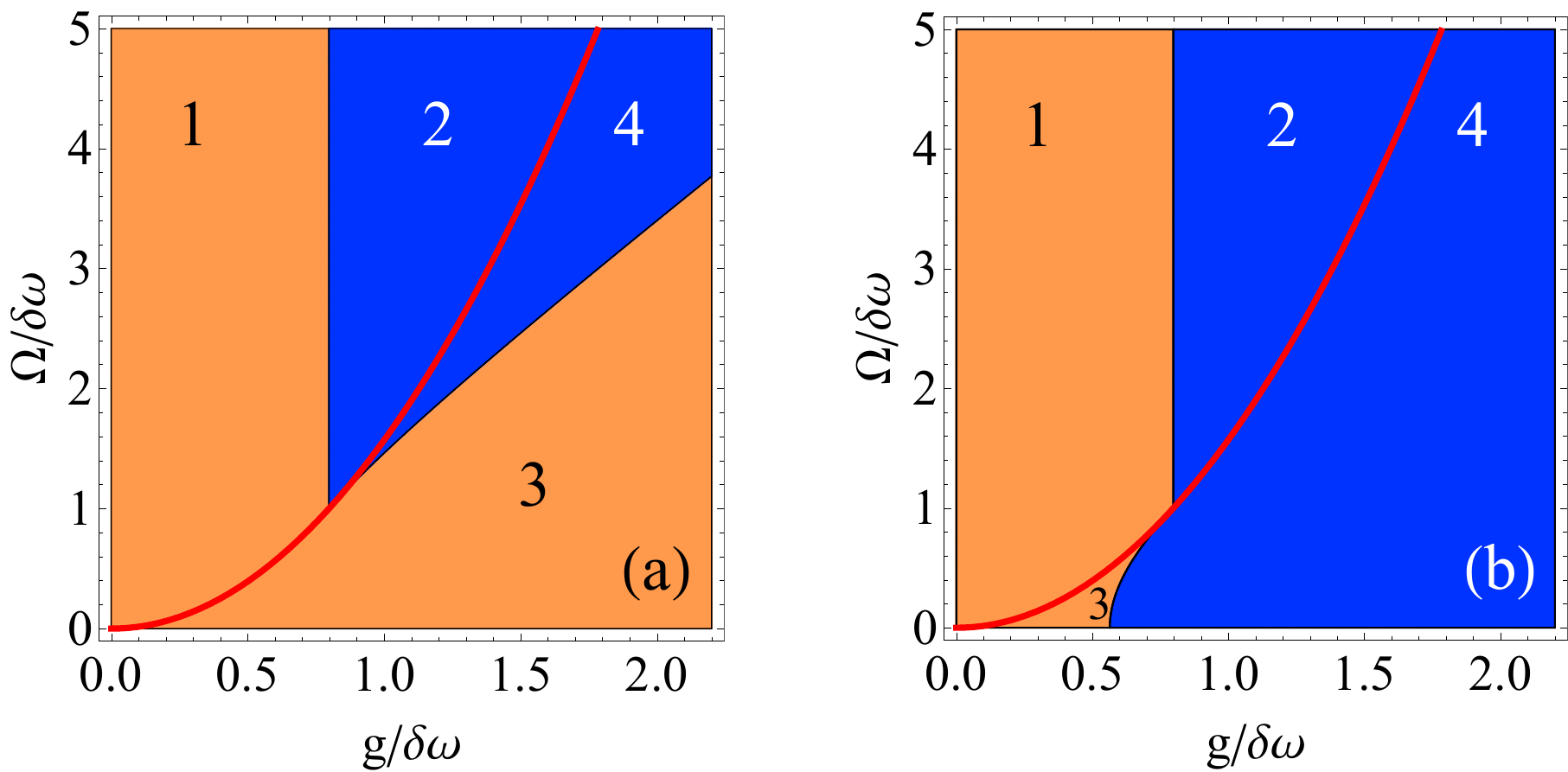}
\caption{Phase diagram of the states in which $\left| a_1\right| \ge \left| a_1\right|$ (a) and the state in which  $\left| a_1\right| \le \left| a_1\right|$ (b). The orange color indicates areas where states are loss-protected by non-Markovian effects. The blue color indicates areas where all states are unprotected. The numbers indicate areas whose boundaries are determined by the conditions  discussed in the section \textit{the influence of PT-symmetry transition on the time of life of states}. The boundaries of first and second areas are determined by conditions~$\Omega>\Omega_{EP}$ (see (\ref{eq:11})) and~(\ref{eq:12}) and~(\ref{eq:13}, respectively. The boundaries of third and fourth areas are determined by conditions~$\Omega<\Omega_{EP}$, and~(\ref{eq:14}), (\ref{eq:15}), respectively.}
\label{fig4}
\end{figure}

When $\Omega > \Omega_{EP} = \gamma /2$, the relaxation rates of the eigenstates are the same and regardless of the initial conditions, the amplitudes of the resonator modes decrease exponentially with the damping decrement $\Gamma = \Gamma_{1,2} = \gamma/2$. In this case, the number of environmental modes excited by the energy flow from the resonators, $N_{ex}$, is proportional to $\gamma / \left(2 \delta \omega \right)= \pi g^2 /\left( \delta \omega ^2 \right)$. Using the fact that the initial state is infinitely living when the number of excited modes $N_{ex} \lesssim 1$, we obtain the condition when the state is loss-protected: $\gamma / \delta\omega \lesssim 1$. Taking into account the expression for $\gamma$, we obtain that both states are loss-protected when
\begin{equation}
\frac{\Omega}{\delta\omega} > \frac{\pi}{2} \left(\frac{g}{\delta\omega}\right)^2
\label{eq:11}
\end{equation}
and 
\begin{equation}
g < \sqrt{\frac{2}{\pi}} \delta\omega
\label{eq:12}
\end{equation}
At the same time, both states are unprotected when 
\begin{equation}
g > \sqrt{\frac{2}{\pi}} \delta\omega
\label{eq:13}
\end{equation}

When $\Omega < \Omega_{EP} = \gamma /2$, the eigenstates have a non-symmetric distribution between resonators and different relaxation rates. As a result, the eigenstates have different areas, in which they are loss-protected. Using the expression~(\ref{eq:7}) and the condition $N_{ex} \lesssim 1$, we obtain that the initial state, which has a maximum in a resonator that does not interact with the environment is loss-protected when
\begin{equation}
\frac{\pi g^2}{2 \delta\omega^2}-\frac{1}{2} \sqrt{\frac{\pi^2 g^4}{\delta\omega^4} - \frac{4 \Omega^2}{\delta\omega^2}} < 1
\label{eq:14}
\end{equation}
When $\Omega << \Omega_{EP}$ this condition takes the form $\Omega < g$.

The initial state, which has a maximum in a resonator that interacts with the environment is loss-protected when
\begin{equation}
\frac{\pi g^2}{2 \delta\omega^2}+\frac{1}{2} \sqrt{\frac{\pi^2 g^4}{\delta\omega^4} - \frac{4 \Omega^2}{\delta\omega^2}} <1
\label{eq:15}
\end{equation}
When $\Omega << \Omega_{EP}$ this condition takes the form $g < \delta\omega / \sqrt{\pi} $.

Using the obtained conditions, we construct a phase diagram of the states [Figure~\ref{fig4}]. It is seen that the obtained estimates of the stability regions predict behavior similar to that obtained in numerical calculations of memory, $M$ [cf. Figures~\ref{fig3} and \ref{fig4}]. For $g/\delta\omega<1$, all states are loss-protected. For $g/\delta\omega>1$ the state in which $\left| a_1\right| > \left| a_1\right|$ is loss-protected when $g> \Omega$ and is not protected when $g< \Omega$. At the same time, the state in which $\left| a_1\right| > \left| a_1\right|$ is not protected for any values of $\Omega$ when $g/\delta\omega>1$. The boundaries of the regions with different numbers of loss-protected states are determined by the relaxation rates obtained from non-Hermitian equations~(\ref{eq:6}). In particular, the difference in the behavior of the states at $\Omega< \Omega_{EP}$ is due to the difference in the real parts of the eigenstates~(\ref{eq:7}) below the exceptional point. Thus, we conclude that spontaneous PT-symmetry breaking can be observed in the system under consideration.

It is important to emphasize that the proposed system can be based on non-dissipative structures. An example of such a system can be two interacting superconducting qubits, one of which is coupled to a superconducting waveguide \cite{9}. In such a structure, the multimode superconducting waveguide plays the role of the finite-size environment with which energy exchange occurs. If the conditions on the coupled strengths between the qubits and the waveguide ($g$ and $\Omega$) are satisfied, the excited state of the qubits will be infinitely long-lived (loss-protected). Thus, the obtained results show that effects associated with PT-symmetry breaking can be observed in non-dissipative systems.
Changing $g$ and $\Omega$, the system switches between modes with two, one, or no protected states that opens the way to efficient control of quantum states.

\textit{Conclusion.} In conclusion, we consider the quasi-PT-symmetric system consisting of two coupled single-mode resonators, one of which interacts with the finite-size environment. We demonstrate that in such a system the non-Markovian effects that are due to the finite-size of environment can make the state of the system infinitely living (loss-protected) even in the absence of gain. There are regions of parameters in which all states of the system are loss-protected, one specific state is protected, or there are no protected states. The boundaries of the regions with different numbers of loss-protected states are determined by the relaxation rates in the quasi-PT-symmetric system. In particular, the boundaries of the regions depend on the transition at the exceptional point in the quasi-PT-symmetric system. The proposed system can be based on non-dissipative structures, which opens the way to observe effects associated with PT-symmetry breaking in non-dissipative systems.

\section*{Acknowledgments}
The study was financially supported by a Grant from Russian Science Foundation (Project No. 23-42-10010, https://rscf.ru/en/project/23-42-10010/). T.T.S., A.A.Z. and E.S.A thank the theoretical physics and mathematics advancement foundation “Basis”.


\nocite{*}

\bibliography{apssamp}

\end{document}